\documentclass[prd,preprint,superscriptaddress,showpacs,byrevtex]{revtex4}
\usepackage{bm}
\def\fsl#1{\setbox0=\hbox{$#1$}           
   \dimen0=\wd0                                 
   \setbox1=\hbox{/} \dimen1=\wd1               
   \ifdim\dimen0>\dimen1                        
      \rlap{\hbox to \dimen0{\hfil/\hfil}}      
      #1                                        
   \else                                        
      \rlap{\hbox to \dimen1{\hfil$#1$\hfil}}   
      /                                         
   \fi}                                         %
\newcommand{\be}{\begin{equation}}
\newcommand{\ee}{\end{equation}}
\newcommand{\bea}{\begin{eqnarray}}
\newcommand{\eea}{\end{eqnarray}}
\newcommand{\beq}{\begin{equation}}
\newcommand{\eeq}{\end{equation}}
\newcommand{\beqs}{\begin{eqnarray}}
\newcommand{\eeqs}{\end{eqnarray}}

\newcommand{\gsim}{\mathrel{\raisebox{-
.6ex}{$\stackrel{\textstyle>}{\sim}$}}}
\newcommand{\dslash}{D\hspace{-0.067in}\slash}

\begin{document}
\title{ Jet Quenching and Gluon to Hadron Fragmentation Function in Non-Equilibrium QCD at RHIC and LHC }
\author{Gouranga C Nayak }
\affiliation{ 22 West Fourth Street \#1, Lewistown, Pennsylvania 17044, USA }
%
%
\begin{abstract}
Theoretical understanding of the observed jet quenching measurements at RHIC and LHC
is challenging in QCD because it requires understanding of parton to hadron fragmentation
function in non-equilibrium QCD. In this paper, by using closed-time path integral formalism,
we derive the gauge invariant definition of the gluon to hadron fragmentation function
in non-equilibrium QCD which is consistent with factorization theorem in non-equilibrium
QCD from first principles.
\end{abstract}
\pacs{12.38.Mh, 12.38.Aw, 12.38.Lg, 13.87.Fh }
\maketitle
\pagestyle{plain}
\pagenumbering{arabic}
\section{Introduction}
During the early stage of the evolution of the universe, just after $\sim 10^{-12}$ seconds of the big-bang, the universe was
filled with a state of matter known as quark-gluon plasma. This early stage of the universe is known as the
quark epoch. The temperature
of the quark-gluon plasma is very high $\gsim 3.2 \times 10^{12}$ K ($\gsim $ 200 MeV) which is about million times larger than
the temperature of the sun ($\sim 1.56 \times 10^{7}$ K). Besides black holes, the quark-gluon
plasma is much denser than all the forms of the matter we know so far such as neutron stars etc.. Hence it will
be a huge step for us if we can recreate this early universe scenario in the laboratory, {\it i. e.}, if we can
produce quark-gluon plasma in the laboratory. At present, the RHIC and LHC
heavy-ion colliders are the best facilities to create quark-gluon plasma in the laboratory.

RHIC (relativistic heavy ion collider) at BNL collides two gold
nuclei at $\sqrt{s}$ = 200 GeV and the LHC (large hadron collider) at CERN (in its first run) collides two lead
nuclei at $\sqrt{s}$ = 2.76 TeV. Since the total energy at RHIC and LHC are very high $\sim$ 40 TeV and $\sim$ 574 TeV
respectively which are deposited in small volume at the initial time,
the initial energy density might have been much higher than the equivalent temperature of $\sim $ 200 MeV
to produce quark-gluon plasma at RHIC and LHC. In the second run at the LHC in the PbPb collisions
at $\sqrt{s}$ = 5.5 TeV the total energy $\sim$ 1150 TeV will even create much
higher initial energy density.

However, the main problem at RHIC and LHC is whether the quark-gluon plasma is thermalized or not.
In case of a non-thermalized quark-gluon plasma at RHIC and LHC the notion of temperature does not exist and hence
the finite temperature lattice QCD studies become inapplicable at RHIC and LHC. The question of
non-equilibrium quark-gluon plasma arises because the two nuclei at RHIC and LHC travel almost at speed of
light. This means the partons inside the two nuclei just before the collision at RHIC and LHC
carry very high longitudinal momentum and very small transverse momentum creating momentum anisotropy
after the nuclear collision. In order for these anisotropic partons to form thermalized quark-gluon
plasma many more secondary partonic collisions over a large period of time is required. However,
the typical hadronization time in QCD is very small ($\sim 10^{-24}$ seconds) after which the partons
become hadrons. Hence there may not be enough time for these anisotropic partons to thermalize
before hadronization takes place at RHIC and LHC. Hence it is necessary to study nonequilibrium-nonperturbative QCD
at RHIC and LHC which, of course, is a difficult problem.

It should be remembered that the quark-gluon plasma can not be directly detected at RHIC and LHC because
quarks and gluons are not directly observed. The primary indirect signatures of quark-gluon plasma detection are, 1) $j/\psi$
suppression \cite{satz}, 2) dilepton and direct photon production \cite{dilepton}, 3) strangeness enhancement
\cite{strange} and, 4) jet quenching \cite{jetq,marco} etc..
We will focus on the jet fragmentation in non-equilibrium QCD in this paper.

Non-equilibrium quantum field theory can be studied by using closed-time path (CTP) integral formalism \cite{schw,keldysh}.
Hence in order to study nonequilibrium-nonperturbative QCD it is necessary to use path integral formulation
of non-equilibrium QCD by using CTP formalism. It should be mentioned, however, that implementing CTP in non-equilibrium
at RHIC and LHC is a very difficult problem, especially due to the presence of gluons in non-equilibrium and hadronization etc.
Even the calculation of one-loop gluon self energy in non-equilibrium in covariant gauge which is derived
in \cite{greiner,cooper} becomes much more tedious than the calculation of one loop QCD diagram in vacuum due to the presence
of additional closed time path indices in non-equilibrium.

Recently, by using Schwinger-Keldysh closed-time path integral formalism, we have derived the
gauge invariant definition of the quark to hadron fragmentation function in non-equilibrium QCD \cite{nayakfg,nayaka3}.
In this paper we will extend this study
to derive the gauge invariant definition of the gluon to hadron fragmentation function in non-equilibrium
QCD which is obtained from the single gluon {\it in}-state $|g>$ in non-equilibrium QCD.

The paper is organized as follows. In section II we present the main results and discuss
the novel features. In section III we give a brief description of
closed-time path integral formalism in non-equilibrium QCD. We implement closed-time path
integral formalism in non-equilibrium QCD to derive the definition of the gauge non-invariant gluon to hadron fragmentation
function in non-equilibrium QCD in section IV. In section V we prove factorization of infrared divergences in
non-equilibrium QCD. In section VI we derive the definition of the gauge invariant gluon to hadron
fragmentation function in non-equilibrium QCD. Section VII contains conclusions.

\section{ Main Results and Novel Features }

Before presenting the technical details of the derivation of the gluon to hadron fragmentation
function in non-equilibrium QCD we will first summarize the main result and its novel features in this section.
The readers who are not interested in the technical details can skip the rest of the paper.

\subsection{Main Results}

For the gluon with arbitrary non-equilibrium (non-isotropic) distribution function $f_g({\vec k})$ at initial
time $t_0$, we find that the gauge invariant definition of the gluon to hadron fragmentation function in non-equilibrium QCD
which is obtained from the single gluon {\it in}-state $|g>$ and is consistent with factorization theorem in non-equilibrium
QCD is given by
\bea
&& D_{H/g}(z,P_T)
= \frac{k^+}{16z~[1+f_g(k^+,k_T)]} \int dx^- \frac{d^{d-2}x_T}{(2\pi)^{d-1}}  e^{i{k}^+ x^- + i {P}_T \cdot x_T/z} \nonumber \\
&&<in| Q^{\mu a}(x^-,x_T) \Phi[x^-,x_T]a^\dagger_H(P^+,0_T)  a_H(P^+,0_T)  \Phi[0]Q_\mu^a(0) |in>
\label{gnf}
\eea
which is valid in covariant gauge, in light-cone gauge, in general axial gauges, in general non-covariant gauges and in
general Coulomb gauge etc. where $Q^{\mu a}(x)$ is the (quantum) gluon field which fragments to hadron and $|in>$ is the initial state of the
non-equilibrium QCD medium in the Schwinger-Keldysh $in-in$ closed-time path formalism and the path ordered exponential
\bea
\Phi_{ab}[x ]=\left[{\cal P}~ {\rm exp}[-ig\int_{0}^\infty d\lambda~ l \cdot A^c(x +l \lambda )~T^{(A)c}]\right]_{ab},~~~~~~~~~~~T^{(A)c}_{ab}=-if^{cab}
\label{wilg}
\eea
is the non-abelian phase or non-abelian gauge link in the adjoint representation of SU(3) where $l^\mu$ is the light-like four-velocity
and $A^{\mu c}(x)$ is the SU(3) pure gauge background field with $a,b,c=1,2,...,8$.
The definition of the gluon to hadron fragmentation function
in non-equilibrium QCD in eq. (\ref{gnf}) is gauge invariant with respect to the gauge transformation
\bea
T^cA'^c_\mu(x) = U(x)T^cA^c_\mu(x) U^{-1}(x)+\frac{1}{ig}[\partial_\mu U(x)] U^{-1}(x)
\label{aftgrmpi}
\eea
along with the homogeneous transformation \cite{thooft,zuber,zuber1,abbott}
\bea
T^cQ'^c_\mu(x)=U(x)T^cQ^c_\mu(x) U^{-1}(x)
\label{jprtpi}
\eea
where
\bea
U(x)=e^{igT^c\omega^c(x)}.
\label{ux}
\eea
The special case $f_g(\vec{k})=\frac{1}{e^{\frac{k_0}{T}}- 1}$ corresponds to the finite temperature QCD in equilibrium.

We will present a derivation of eq. (\ref{gnf}) in this paper.

\subsection{Novel and Qualitatively Interesting Features}

Let us describe the novel features of the definition of the gluon to hadron fragmentation function in non-equilibrium QCD
obtained in eq. (\ref{gnf}).

Similar to DGLAP evolution equation \cite{dglap} in pp collisions we will follow the DGLAP-like
evolution equation procedure in AA collisions at RHIC and LHC where the inclusive hadron production cross section is given by
the factorized formula in non-equilibrium QCD
\bea
\frac{d\sigma^H_{AA}}{dy dp_T^2}=\sum_i \int \frac{dz}{z^2} \frac{d{\hat \sigma}_i}{dy dp_{Ti}^2} D_{H/i}(z,Q)
\label{fmed}
\eea
where $\frac{d{\hat \sigma}_i}{dy dp_{Ti}^2}$ is the partonic level differential
cross section in non-equilibrium QCD and $D_{H/i}(z,Q)$ is the fragmentation function in non-equilibrium QCD.
The $Q^2$ evolution of the fragmentation function $D_{H/i}(z,Q)$ in non-equilibrium QCD
obeys the generalized DGLAP-like evolution equation in non-equilibrium QCD
\bea
\frac{dD_{H/i}(z,Q)}{d ln Q} = \frac{\alpha_s(Q)}{\pi} \sum_j \int_z^1 \frac{dz'}{z'} P_{i\rightarrow jk}(z',Q) D_{H/j}(\frac{z}{z'},Q)
\label{evol}
\eea
where $P_{i\rightarrow jk}(z',Q)$ is the splitting function in non-equilibrium QCD \cite{nayakdglap}. In the above
equations the indices $i,j,k=q,{\bar q},g$.

Note that one of the crucial ingredient to derive DGLAP evolution equation is the factorization theorem \cite{dglap}.
Hence similar to DGLAP equation in QCD in vacuum the DGLAP-like equation in non-equilibrium QCD in eq. (\ref{evol})
is consistent with the factorization theorem in non-equilibrium QCD (see also eq. (\ref{fmed}) and appendix A).
This implies that the
partonic level differential cross section $\frac{d{\hat \sigma}_i}{dy dp_{Ti}^2}$ in
eq. (\ref{fmed}) in non-equilibrium QCD and the fragmentation
function $D_{H/i}(z,Q)$ in eq. (\ref{fmed})
in non-equilibrium QCD are studied by using $|in>$ for the ground state in non-equilibrium QCD instead of the usual vacuum state $|0>$
in QCD. It should be remembered that the ground state $|in>$ in non-equilibrium QCD contains both the vacuum part
and the medium part. For example, when the distribution function $f({\vec k})=0$ then $|in>$ becomes $|0>$ and
one reproduces all the equations and quantities of the QCD in vacuum.

The leading order perturbative gluon propagator in non-equilibrium QCD is given by \cite{greiner,cooper}
\bea
G^{\mu \nu}(k)_{rs} = -i[g^{\mu \nu} +  (\alpha -1) \frac{k^\mu k^\nu}{k^2}] ~G^{\rm vac}_{rs}(k) -iT^{\mu \nu}G^{\rm med}_{rs}(k)
\label{gpm}
\eea
which contains the vacuum propagator $G^{\rm vac}_{rs}(k)$ plus the medium propagator $G^{\rm med}_{rs}(k)$ where
\bea
G^{\rm vac}_{rs}(k)=
\left ( \begin{array}{cc}
\frac{1}{k^2+i\epsilon} & -2\pi \delta(k^2)\theta(-k_0) \\
-2\pi \delta(k^2)\theta(k_0) & -\frac{1}{k^2-i\epsilon}
\end{array} \right )
\label{glp}
\eea
and
\bea
G^{\rm med}_{rs}(k)= 2\pi \delta(k^2) f_g(\vec{k})
\left ( \begin{array}{cc}
1 & 1 \\
1 & 1
\end{array} \right ).
\label{gvc}
\eea
In the above equations the closed-time indices $r,s= +,-$ correspond to upper and lower time branch in the closed-time path formalism.

However, since the gluon fragmentation function is a non-perturbative quantity in QCD
it is not possible to decompose the gluon fragmentation function into the vacuum part and
the medium part (see eq. (\ref{gnf})).
Hence one finds that to be consistent with factorization theorem in non-equilibrium QCD, the fragmentation
function in QCD in vacuum can not be used in the DGLAP-like equation in non-equilibrium QCD in eq. (\ref{evol})
to study hadron production from quark-gluon plasma. The fragmentation function in non-equilibrium QCD should be
used in DGLAP-like equation in non-equilibrium QCD in eq. (\ref{evol}) to study hadron production from
quark-gluon plasma by using factorized formula as given by
eq. (\ref{fmed}) where $\frac{d{\hat \sigma}_i}{dy dp_{Ti}^2}$ is the partonic level differential
cross section in non-equilibrium QCD and $D_{H/i}(z,Q)$ is the fragmentation function in non-equilibrium QCD.

\section{ Non-equilibrium QCD Using Closed-Time Path Integral Formalism }

When a system is in non-equilibrium its asymptotic future is quite different from
its initial preparation in the remote past. Hence unlike zero-temperature field
theory or finite temperature field theory in equilibrium there is no simple relation
between asymptotic future state and initial state in the remote past. For this reason
the generic non-equilibrium quantum field theory is very non-trivial. In order to avoid the
asymptotic future state and to deal with only initial state in the remote past one
introduces two time branches in the closed-time path (CTP) formalism \cite{schw,keldysh}.

The gluon to hadron fragmentation function in non-equilibrium QCD in eq. (\ref{gnf})
is a non-perturbative quantity. Hence it is necessary to use path integral
formulation of non-equilibrium QCD in CTP formalism to study its properties.
Before going to gluons in non-equilibrium QCD let us consider the scalar gluons first in the CTP formalism
in non-equilibrium in the path integral formulation. The generating functional in scalar field theory in
non-equilibrium is given by
\bea
Z[\rho,J_+,J_-]=\int [d\phi_+][d\phi_-] ~{\rm exp}[i\int d^4x [{\cal L}[\phi_+]-{\cal L}[\phi_-]+ J_+\phi_+- J_-\phi_-]]~<\phi_+,0|\rho|0,\phi_-> \nonumber \\
\label{rho7}
\eea
where $\rho$ is the initial density of states in non-equilibrium,
${\cal L}[\phi]$ is the full lagrangian density in scaler field theory and $|\phi_{\pm},0>$ is the quantum state
corresponding to the field configuration $\phi_{\pm}(\vec{x},t=0)$ at the initial time. The subscript $+$ refers to
closed-time path index in the positive time contour and the subscript $-$ refers to
closed-time path index in the negative time contour in the closed-time path formalism \cite{schw,keldysh}.

Since there are two time branches $+,-$ in the closed-time path formalism there are two external sources $J_+,J_-$ corresponding
to these two different time branches. Hence one finds that there are four green's functions in non-equilibrium which are given by
\bea
&& G_{++}(x,x') = \frac{\delta Z[J_+,J_-,\rho]}{i^2 \delta J_+(x) J_+(x')}= <T\phi (x) \phi (x')>=<in|T\phi (x) \phi (x')|in> \nonumber \\
&& G_{--}(x,x') = \frac{\delta Z[J_+,J_-,\rho]}{(-i)^2 \delta J_-(x) J_-(x')} = <{\bar T} \phi (x) \phi (x')>= <in|{\bar T} \phi (x) \phi (x')|in>\nonumber \\
&& G_{+-}(x,x') = \frac{\delta Z[J_+,J_-,\rho]}{-i^2 \delta J_+(x) J_-(x')} = <\phi (x') \phi (x) >= <in|\phi (x') \phi (x)|in>\nonumber \\
&& G_{-+}(x,x') = \frac{\delta Z[J_+,J_-,\rho]}{-i^2 \delta J_-(x) J_+(x')}= <\phi (x) \phi (x') >= <in|\phi (x) \phi (x')|in>
\label{green}
\eea
where $|in>$ is the initial state in non-equilibrium quantum field theory at the initial time $t=t_{in}=0$.
Note that due to the presence of the medium the initial state $|in>$ in non-equilibrium
quantum field theory is different from the vacuum state $|0>$ in the quantum field
theory in vacuum.

In order to extend scalar gluon to gluon in non-equilibrium QCD we need to add gauge fixing term
in the lagrangian density which results in the appearance of the Faddeev-Popov (F-P) determinant
in the generating functional which can be expressed in terms of path integral over the ghost fields
\cite{muta}. However, we will directly work with the Faddeev-Popov (F-P) determinant in this paper
and we will work in the frozen ghost formalism \cite{greiner,cooper} for the medium part at the initial time $t=t_{in}=0$.
Hence it is now straightforward to extend the generating functional of the scalar gluon in non-equilibrium in eq. (\ref{rho7})
to gluon in non-equilibrium QCD.

Extending the generating functional of the scalar gluon in non-equilibrium in eq. (\ref{rho7})
to gluon in non-equilibrium QCD we find that the generating functional in non-equilibrium QCD
in the path integral formulation is given by \cite{greiner,cooper}
\bea
&& Z[\rho,J_+,J_-,\eta_+,\eta_-,{\bar \eta}_+,{\bar \eta}_-]=\int [dQ_+] [dQ_-][d{\bar \psi}_+] [d{\bar \psi}_-] [d \psi_+ ] [d\psi_-]~
{\rm det}(\frac{\delta \partial_\mu Q_+^{\mu c}}{\delta \omega_+^d}){\rm det}(\frac{\delta \partial_\mu Q_-^{\mu c}}{\delta \omega_-^d}) \nonumber \\
&& {\rm exp}[i\int d^4x [-\frac{1}{4}({F^c}_{\mu \nu}^2[Q_+]-{F^c}_{\mu \nu}^2[Q_-])-\frac{1}{2 \alpha}
(\partial_\mu Q_+^{\mu c })^2+\frac{1}{2 \alpha}(\partial_\mu Q_-^{\mu c })^2 +{\bar \psi}_+ \dslash [Q_+] \psi_+ \nonumber \\
&& -{\bar \psi}_- \dslash [Q_-] \psi_-
+ J_+ \cdot Q_+ -J_- \cdot Q_-+{\bar \eta}_+ \cdot \psi_+ - {\bar \eta}_- \cdot \psi_- +
 {\bar \psi}_+ \cdot \eta_+-  {\bar \psi}_- \cdot \eta_- ]] \nonumber \\
&& \times ~<Q_+,\psi_+,{\bar \psi}_+,0|~\rho~|0,{\bar \psi}_-,\psi_-,Q_->
\label{zfqinon}
\eea
where $\rho$ is the initial density of state in non-equilibrium QCD, the state $|Q^\pm,\psi^\pm,{\bar \psi}^\pm,0>$ corresponds to the field
configurations $Q_\mu^c({\vec x},t=t_{in}=0)$, $\psi({\vec x},t=t_{in}=0)$, ${\bar \psi}({\vec x},t=t_{in}=0)$, the
$J^{\mu c}(x)$ is the external source for the quantum gluon field $Q^{\mu c}(x)$, the ${\bar \eta}_i(x)$ is the external source for the
Dirac field $\psi_i(x)$ of the quark, $\alpha$ is the gauge fixing parameter, ${\rm det}(\frac{\delta (\partial_\mu Q^{\mu b}_\pm)}{\delta \omega^c_\pm})$ is the Faddeev-Popov (F-P) determinant and
\bea
&&F^c_{\mu \nu}[Q_\pm]=\partial_\mu Q^c_{\nu \pm}(x)-\partial_\nu Q^c_{\mu \pm}(x)+gf^{cba}Q^b_{\mu \pm}(x)Q^a_{\nu \pm}(x),~~~~~~~~~{F^c}_{\mu \nu}^2[Q_\pm]={F}^{\mu \nu c}[Q_\pm]{F}^c_{\mu \nu}[Q_\pm], \nonumber \\
&& \dslash[Q_\pm] =i\gamma^\mu \partial_\mu +gT^c\gamma^\mu Q^c_{\mu \pm}.
\eea
As mentioned above the Faddeev-Popov (F-P) determinant ${\rm det}(\frac{\delta (\partial_\mu Q^{\mu b}_\pm)}{\delta \omega^c_\pm})$
in eq. (\ref{zfqinon}) can be expressed in terms of path integral over the ghost fields \cite{muta} but we will directly work with
the Faddeev-Popov (F-P) determinant ${\rm det}(\frac{\delta (\partial_\mu Q^{\mu b}_\pm)}{\delta \omega^c_\pm})$ in this paper.

Note that the generating functional in non-equilibrium QCD in eq. (\ref{zfqinon})
is generic from which the finite temperature QCD formalism and QCD in vacuum formalism can be obtained.

From eq. (\ref{zfqinon}) we find that the non-perturbative gluon correlation function in non-equilibrium QCD is given by
\bea
&& <in|Q_{\mu r}^a(x_1)Q_{\nu s}^b(x_2)|in>=\int [dQ_+] [dQ_-][d{\bar \psi}_+] [d{\bar \psi}_-] [d \psi_+ ] [d\psi_-]~
Q_{\mu r}^a(x_1)Q_{\nu s}^b(x_2) \nonumber \\
&& {\rm det}(\frac{\delta \partial_\mu Q_+^{\mu c}}{\delta \omega_+^d}){\rm det}(\frac{\delta \partial_\mu Q_-^{\mu c}}{\delta \omega_-^d}) {\rm exp}[i\int d^4x [-\frac{1}{4}{F^c}_{\mu \nu}^2[Q_+]+\frac{1}{4}{F^c}_{\mu \nu}^2[Q_-]-\frac{1}{2 \alpha}
(\partial_\mu Q_+^{\mu c })^2+\frac{1}{2 \alpha} (\partial_\mu Q_-^{\mu c })^2\nonumber \\
&& +{\bar \psi}_+ \dslash [Q_+] \psi_+ -{\bar \psi}_- \dslash [Q_-] \psi_-]]<Q_+,\psi_+,{\bar \psi}_+,0|~\rho~|0,{\bar \psi}_-,\psi_-,Q_->
\label{nezf}
\eea
where $r,s=+,-$ are the closed-time path indices ($"+"$ sign corresponds to positive time path and $"-"$ sign
corresponds to negative time path in the closed-time path formalism \cite{schw,keldysh}).

\section{ Gauge Non-Invariant Definition of Gluon to Hadron Fragmentation Function in Non-Equilibrium QCD }

For simplicity, let us consider the scalar gluon in non-equilibrium first before considering
the gluon in non-equilibrium QCD. For the application to RHIC and LHC high energy nuclear collisions
it is useful to consider the light-cone coordinate system $x^\mu=(x^+,x^-,x_T)$.
Note that at the initial time $x^+=x^+_{in}=0$, the interaction picture coincides with the Heisenberg picture
and Schrodinger picture. Hence at the initial time $x^+=x^+_{in}=0$ we write the scalar field as
\bea
\phi(x^-,x_T)=\frac{1}{(2\pi)^{d-1}} \int \frac{dp^+}{\sqrt{2p^+}}d^{d-2}p_T [e^{-ip^+x^-+ip_T \cdot x_T} a(p^+,p_T)+e^{ip^+x^--ip_T \cdot x_T} a^\dagger(p^+,p_T)]\nonumber \\
\label{phi}
\eea
where $a^\dagger(p)$ and $a(p)$ are the creation and the annihilation operators of the scalar gluon respectively.
The commutation relations at the initial time are given by
\bea
&& [a(p^+,p_T), a^\dagger ({p^\prime}^+, p^\prime_T) ]_{x^+=0} = (2\pi)^{d-1} \delta(p^+-{p^\prime}^+) \delta^{(d-2)} (p_T - p^\prime_T), \nonumber \\
&& [a(p^+,p_T), a ({p^\prime}^+,p^\prime_T) ]_{x^+=0} =[a^\dagger( p^+,p_T), a^\dagger ({ p^\prime }^+, p'_T) ]_{x^+=0} =0.
\label{comml}
\eea
The non-equilibrium distribution function
$f(p^+,p_T)$ of the fragmenting scalar gluon at initial time is given by
\bea
<in|a^\dagger (p^+,p_T)a({p^\prime}^+,p^\prime_T)|in> = f(p^+,p_T) (2\pi)^{d-1} \delta(p^+-{p^\prime}^+) \delta^{(d-2)} ( p_T - p^\prime_T)
\label{distl}
\eea
where we have assumed the space ($x^-$ and $x_T$) translational invariance at initial time $x^+=x^+_{in}=0$.
The special case $f_g(\vec{p})=\frac{1}{e^{\frac{p_0}{T}}- 1}$ corresponds to the finite temperature field theory in equilibrium.

At the initial time  $x^+=x^+_{in}=0$ the scalar gluon state in the non-equilibrium quantum field theory is given by \cite{nayakfg}
\bea
|p^+, p_T> = a^\dagger (p^+, p_T) |in>
\label{kt}
\eea
where $a^\dagger(p)$ is the creation operator of the scalar gluon and $|in>$ is the initial state of the non-equilibrium medium.
By using eqs. (\ref{kt}), (\ref{distl}) and (\ref{comml}) we find, at initial time,
\bea
&& <p^+, p_T|{p'}^+, {p'}_T> = <in|a(p^+, p_T)a^\dagger({p'}^+, {p'}_T)|in> \nonumber \\
&& =(2\pi)^{d-1} \delta(p^+-{p'}^+) \delta^{d-2}(p_T -{p'}_T) [1+f(p^+,p_T)].
\label{norml}
\eea

Consider the inclusive production of hadron $H$ created in the $out-$state $|H+X>$ from
a scalar gluon in non-equilibrium in the initial state $|p>$ with the probability amplitude
\bea
<H+X|p>,
\label{amp}
\eea
where $X$ being other outgoing final state hadrons.
The distribution $h_p(P)$ of the hadron $H$ with momentum $P$ from the parton of
momentum $p$ in non-equilibrium can be found from the amplitude in eq. (\ref{amp}) by using
\bea
\sum_X~<p,p_T|H+X><H+X|{p^+}',p'_T>=h_p(P) <p^+,p_T|{p^+}',p'_T>.
\label{hp1l}
\eea
We write the left hand side as
\bea
&& \sum_X~<p^+,p_T|H+X><H+X|{p^+}',p'_T>= \sum_X~<p^+,p_T|a^\dagger_H(P)|X><X|a_H(P)|{p^+}',p'_T> \nonumber \\
&& =<p^+,p_T|a^\dagger_H(P)a_H(P)|{p^+}',p'_T>.
\label{hp2l}
\eea
From eqs. (\ref{hp1l}), (\ref{hp2l}), (\ref{kt}) and (\ref{norml}) we find
\bea
&&<in|a(p^+,p_T)a^\dagger_H(P^+,P_T)a_H(P^+,P_T)a^\dagger({p'}^+,{p'}_T)|in> \nonumber \\
&&=2z(2\pi)^{d-1}D_{H/a}(z,P_T)~[1+f(p^+,p_T)]\delta(p^+-{p'}^+) \delta^{d-2}(p_T -{p'}_T).
\label{fr1lal}
\eea
From eq. (\ref{phi}) we obtain
\bea
&& (2\pi)^{d-1} <in| \phi(x^-,x_T) a^\dagger_H(P^+,P_T)  a_H(P^+,P_T) \phi(0) |in>
 = \frac{1}{(2\pi)^{d-1}} \int \frac{dp^+}{\sqrt{2p^+}} d^{d-2}p_T \int \frac{d{p'}^+}{\sqrt{2{p'}^+}} d^{d-2}{p'}_T \nonumber \\
  &&~[<in| e^{-ip\cdot x} a(p^+,p_T)a^\dagger_H(P^+,P_T)a_H(P^+,P_T)a^\dagger({p'}^+,{p'}_T)|in>]_{x^+=0}.
\label{fr3m}
\eea

Using this in eq. (\ref{fr1lal}) we find the expression of the scalar gluon fragmentation function in non-equilibrium
\bea
&& D_{H/a}(z,P_T)= \frac{p^+}{z~[1+f(p^+,p_T)]} \int dx^- \frac{d^{d-2}x_T}{(2\pi)^{d-1}}  e^{i{p}^+ x^- - i {p}_T \cdot x_T} \nonumber \\
&& <in| \phi(x^-,x_T) a^\dagger_H(P^+,P_T)  a_H(P^+,P_T) \phi(0) |in>.
\label{fr6ollp}
\eea
By making a Lorentz transformation to the zero transverse momentum frame of the hadron \cite{nayakfg} we find from eq. (\ref{fr6ollp})
that the scalar gluon fragmentation function in non-equilibrium is given by
\bea
&& D_{H/a}(z,P_T)= \frac{p^+}{z~[1+f(p^+,p_T)]} \int dx^- \frac{d^{d-2}x_T}{(2\pi)^{d-1}}  e^{i{p}^+ x^- + i {P}_T \cdot x_T/z} \nonumber \\
&& <in| \phi(x^-,x_T) a^\dagger_H(P^+,0_T)  a_H(P^+,0_T) \phi(0) |in>.
\label{sgl}
\eea
In the above expression, $f(p^+,p_T)$ is the non-equilibrium distribution function of the fragmenting scalar gluon
at initial time $x^+=x^+_{in}=0$ and $|in>$ is the initial state of the non-equilibrium medium in the
Schwinger-Keldysh $in-in$ closed-time path formalism.

In analogous to eq. (\ref{sgl}) for the scalar gluon case, we find the gauge non-invariant definition of the
gluon to hadron fragmentation function in non-equilibrium QCD is given by
\bea
&& D_{H/g}(z,P_T)
= \frac{p^+}{16z~[1+f_g(p^+,p_T)]} \int dx^- \frac{d^{d-2}x_T}{(2\pi)^{d-1}}  e^{i{p}^+ x^- + i {P}_T \cdot x_T/z} \nonumber \\
&&<in| Q^{\mu c}(x^-,x_T) a^\dagger_H(P^+,0_T)  a_H(P^+,0_T)  Q_\mu^c(0) |in>
\label{gnon}
\eea
where $f_g(p^+,p_T)$ is the non-equilibrium distribution function of the fragmenting gluon in QCD at initial time,
$Q^{\mu c}(x)$ is the (quantum) gluon field in non-equilibrium QCD which fragments to hadron $H$
and $|in>$ is the initial state of the non-equilibrium QCD medium in the Schwinger-Keldysh $in-in$ closed-time path formalism.

Note that the transition from massless scalar gluon fragmentation function
in eq. (\ref{sgl}) to gauge non-invariant gluon fragmentation function in QCD in
eq. (\ref{gnon}) is achieved by the following procedure.
The massless scalar gluon fragmentation function in quantum field theory in eq. (\ref{sgl}) is proportional to the
two-point scalar gluon correlation function of the type
$<in|\phi(x) a^\dagger_H a_H \phi(0)|in>$. Note that, in quantum field theory, the two-point scalar gluon correlation
function in scalar field theory is given by $<in|\phi(x) \phi(0)|in>$ and the two-point gluon
correlation function in QCD is given by $<in|Q_\mu^a(x) Q_\nu^b(0)|in>$ where $\mu, \nu$ are
Lorentz indices and $a, b$ are color indices of the gluon fields $Q_\mu^a(x)$ and $Q_\nu^b(0)$ in QCD.
As mentioned above, in quantum field theory, the gluon fragmentation function is proportional to two-point
gluon correlation function.
Hence the gluon fragmentation function in QCD in eq. (\ref{gnon}) is obtained from the massless scalar gluon
fragmentation function in eq. (\ref{sgl}) by replacing
\bea
<in|\phi(x) a^\dagger_H a_H \phi(0)|in> ~~~\rightarrow ~~~ <in|Q_\mu^a(x) a^\dagger_H a_H Q^{\mu a}(0)|in>.
\eea
Since the fragmentation function in QCD is a Lorentz scalar and color neutral quantity,
the sum over Lorentz and color indices are taken in eq. (\ref{gnon}).
The additional factor $\frac{1}{16}$ in eq. (\ref{gnon})
is supplied because the average over the spin and color ($\frac{1}{2 \times 8}$) of the gluon in QCD
is taken which is absent in the scalar gluon case.

\section{ Proof of Factorization of Infrared Divergences in Non-Equilibrium QCD }

Since the gluon to hadron fragmentation function in non-equilibrium QCD in eq. (\ref{gnon}) is a non-perturbative
quantity we will use path integral formulation of quantum field theory for this purpose \cite{tucci,nayakarx}.

The generating functional in the background field method of QCD in
non-equilibrium QCD in the path integral formulation is given by \cite{greiner,cooper,thooft,abbott,zuber,zuber1}
\bea
&& Z[\rho,A,J_+,J_-,\eta_+,\eta_-,{\bar \eta}_+,{\bar \eta}_-]=\int [dQ_+] [dQ_-][d{\bar \psi}_+] [d{\bar \psi}_-] [d \psi_+ ] [d\psi_-]~
 \nonumber \\
&& {\rm det}(\frac{\delta G^c(Q_+)}{\delta \omega_+^d})~\times ~{\rm det}(\frac{\delta G^c(Q_-)}{\delta \omega_-^d}) \times {\rm exp}[i\int d^4x [-\frac{1}{4}{F^c}_{\mu \nu}^2[Q_++A_+]+\frac{1}{4}{F^c}_{\mu \nu}^2[Q_-+A_-]\nonumber \\
&&-\frac{1}{2 \alpha} (G^c(Q_+))^2+\frac{1}{2 \alpha}(G^c(Q_-))^2 +{\bar \psi}_+ \dslash [Q_++A_+] \psi_+ \nonumber \\
&& -{\bar \psi}_- \dslash [Q_-+A_-] \psi_-
+ J_+ \cdot Q_+ -J_- \cdot Q_-+{\bar \eta}_+ \cdot \psi_+ - {\bar \eta}_- \cdot \psi_- + {\bar \psi}_+ \cdot \eta_+
-  {\bar \psi}_- \cdot \eta_- ]]
\nonumber \\
&& <Q_++A_+,\psi_+,{\bar \psi}_+,0|~\rho~|0,{\bar \psi}_-,\psi_-,Q_-+A_->
\label{azaqcd}
\eea
where the gauge fixing term is given by
\bea
G^c(Q_\pm) =\partial_\mu Q^{\mu c}_\pm + gf^{cba} A_{\mu \pm}^b Q^{\mu a}_\pm=D_\mu[A_\pm]Q^{\mu c}_\pm
\label{ga}
\eea
which depends on the background field $A^{\mu c}(x)$ and
\bea
F_{\mu \nu}^c[A_\pm+Q_\pm]=\partial_\mu [A_{\nu \pm}^c+Q_{\nu \pm}^c]-\partial_\nu [A_{\mu \pm}^c+Q_{\mu \pm}^c]+gf^{cba} [A_{\mu \pm}^b+Q_{\mu \pm}^b][A_{\nu \pm}^a+Q_{\nu \pm}^a]
\label{fmn}
\eea
where $Q^{\mu c}$ is the quantum gluon field and $A^{\mu c}$ is the background field \cite{thooft,zuber,abbott}.

From eq. (\ref{azaqcd}) we find that the nonequilibrium-nonperturbative gluon correlation function
in the background field method of QCD is given by
\bea
&& <in|Q_{\mu r}^a(x_1)Q_{\nu s}^b(x_2)|in>_A=\int [dQ_+] [dQ_-][d{\bar \psi}_+] [d{\bar \psi}_-] [d \psi_+ ] [d\psi_-]~
Q_{\mu r}^a(x_1)Q_{\nu s}^b(x_2) \nonumber \\
&& {\rm det}(\frac{\delta G^c(Q_+)}{\delta \omega_+^d}){\rm det}(\frac{\delta G^c(Q_-)}{\delta \omega_-^d}) {\rm exp}[i\int d^4x [-\frac{1}{4}{F^c}_{\mu \nu}^2[Q_++A_+]+\frac{1}{4}{F^c}_{\mu \nu}^2[Q_-+A_-]\nonumber \\
&&-\frac{1}{2 \alpha} (G^c(Q_+))^2+\frac{1}{2 \alpha}(G^c(Q_-))^2 +{\bar \psi}_+ \dslash [Q_++A_+] \psi_+ \nonumber \\
&& -{\bar \psi}_- \dslash [Q_-+A_-] \psi_-]]~<Q_++A_+,\psi_+,{\bar \psi}_+,0|~\rho~|0,{\bar \psi}_-,\psi_-,Q_-+A_->.
\label{cfqcd}
\eea

For a light-like quark attached to infinite number of gluons we find the eikonal factor \cite{nayakarx}
\bea
&& 1+ gT^c\int \frac{d^4k}{(2\pi)^4} \frac{l\cdot { A}^c(k)}{l\cdot k +i\epsilon }
+g^2\int \frac{d^4k_1}{(2\pi)^4} \frac{d^4k_2}{(2\pi)^4} \frac{T^c l\cdot { A}^c(k_1)T^b l\cdot { A}^b(k_2)}{(l\cdot k_1 +i \epsilon)(l\cdot (k_1+k_2) +i \epsilon)}+... \nonumber \\
&&={\cal P}~{\rm exp}[ig \int_0^{\infty} d\lambda l\cdot { A}^c(l\lambda)T^c ]
\label{iiij}
\eea
which describes the infrared divergences arising from the infinite number of soft gluons exchange
with the light-like quark of four-velocity $l^\mu$ where ${\cal P}$ is  the path ordering and the
gluon field $ { A}^{\mu c}(x)$ and its Fourier transform $ { A}^{\mu c}(k)$ are related by
\bea
{ A}^{\mu c}(x) =\int \frac{d^4k}{(2\pi)^4} { A}^{\mu c}(k) e^{ik \cdot x}.
\label{ft}
\eea
The light-like quark traveling with light-like four-velocity $l^\mu$ produces SU(3) pure gauge field $A^{\mu c}(x)$
both in classical mechanics  \cite{collinssterman,nayakj,nayake} and in quantum field theory \cite{nayakjp}
at all the time-space position $x^\mu$ except at the position ${\vec x}$ perpendicular to the direction of motion
of the quark (${\vec l}\cdot {\vec x}=0$) at the time of closest approach ($x_0=0$).
When $A^{\mu c}(x) = A^{\mu c}(\lambda l)$ as in eq. (\ref{iiij})
we find ${\vec l}\cdot {\vec x}=\lambda {\vec l}\cdot {\vec l}=\lambda\neq 0$ which implies that the light-like quark
finds the gluon field $A^{\mu c}(x)$ in eq. (\ref{iiij}) as the SU(3) pure gauge. The SU(3) pure gauge is given by
\bea
T^a_{ij}A^{\mu a}(x) =\frac{1}{ig} [[\partial^\mu \Phi(x)]\Phi^{-1}(x)]_{ij}
\label{gtqcd}
\eea
which gives the non-abelian gauge link \cite{nayakarx}
\bea
\Phi_{ij}(x)=[{\cal P}{\rm exp}[-ig\int_0^{\infty} d\lambda l\cdot { A}^c(x+l\lambda)T^{(A)c}]]_{ij}=[e^{igT^a\omega^a(x)}]_{ij}.
\label{wilt}
\eea
The gauge fixing term $\frac{1}{2 \alpha} (G^a(Q))^2$ in eq. (\ref{azaqcd}) [where $G^a(Q)$ is given by eq. (\ref{ga})]
is invariant for gauge transformation of $A_\mu^a$:
\bea
\delta A_\mu^a = gf^{abc}A_\mu^b\omega^c + \partial_\mu \omega^a
\label{typeI}
\eea
provided one also performs a homogeneous transformation of $Q_\mu^a$ \cite{zuber,abbott}:
\bea
\delta Q_\mu^a =gf^{abc}Q_\mu^b\omega^c.
\label{omega}
\eea
By changing the integration variable $Q \rightarrow Q-A$ in the right hand side of eq. (\ref{cfqcd}) we find
\bea
&& <in|Q_{\mu r}^a(x_1)Q_{\nu s}^b(x_2)|in>_A=\int [dQ_+] [dQ_-][d{\bar \psi}_+] [d{\bar \psi}_-] [d \psi_+ ] [d\psi_-]~
[Q^a_{\mu r}(x_1)-A^a_{\mu r}(x_1)]\nonumber \\
&&[Q^b_{\nu s}(x_2)-A^b_{\nu s}(x_2] {\rm det}(\frac{\delta G^c_f(Q_+)}{\delta \omega_+^d}){\rm det}(\frac{\delta G^c_f(Q_-)}{\delta \omega_-^d}) {\rm exp}[i\int d^4x [-\frac{1}{4}{F^c}_{\mu \nu}^2[Q_+]+\frac{1}{4}{F^c}_{\mu \nu}^2[Q_-]\nonumber \\
&&-\frac{1}{2 \alpha} (G^c_f(Q_+))^2+\frac{1}{2 \alpha}(G^c_f(Q_-))^2 +{\bar \psi}_+ \dslash [Q_+] \psi_+ -{\bar \psi}_- \dslash [Q_-] \psi_-]]<Q_+,\psi_+,{\bar \psi}_+,0|~\rho~|0,{\bar \psi}_-,\psi_-,Q_->\nonumber \\
\label{cfqcd1}
\eea
where from eq. (\ref{ga}) we find
\bea
G_f^c(Q_\pm) =\partial_\mu Q^{\mu a}_\pm + gf^{cba} A_{\mu \pm}^b Q^{\mu a}_\pm - \partial_\mu A^{\mu c}_\pm=D_\mu[A_\pm] Q^{\mu c}_\pm - \partial_\mu A^{\mu c}_\pm
\label{gfa}
\eea
and from eq. (\ref{omega}) [by using eq. (\ref{typeI})] we find
\bea
\delta Q_{\mu \pm}^c = -gf^{cba}\omega^b_\pm Q_{\mu \pm}^a + \partial_\mu \omega^c_\pm.
\label{theta}
\eea

Changing the integration variable from unprimed variable to primed variable we find from eq. (\ref{cfqcd1})
\bea
&& <in|Q_{\mu r}^a(x_1)Q_{\nu s}^b(x_2)|in>_A=\int [dQ'_+] [dQ'_-][d{\bar \psi}'_+] [d{\bar \psi}'_-] [d \psi'_+ ] [d\psi'_-]~
[Q'^a_{\mu r}(x_1)-A_{\mu r}^a(x_1)]\nonumber \\
&&[Q'^b_{\nu s}(x_2)-A_{\nu s}^b(x_2]  \times
{\rm det}(\frac{\delta G^c_f(Q'_+)}{\delta \omega_+^d}){\rm det}(\frac{\delta G^c_f(Q'_-)}{\delta \omega_-^d}) {\rm exp}[i\int d^4x [-\frac{1}{4}{F^c}_{\mu \nu}^2[Q'_+]+\frac{1}{4}{F^c}_{\mu \nu}^2[Q'_-]\nonumber \\
&&-\frac{1}{2 \alpha} (G^c_f(Q'_+))^2+\frac{1}{2 \alpha}(G^c_f(Q'_-))^2 +{\bar \psi}'_+ \dslash [Q'_+] \psi'_+ -{\bar \psi}'_- \dslash [Q'_-] \psi'_-]]<Q'_+,\psi'_+,{\bar \psi}'_+,0|~\rho~|0,{\bar \psi}'_-,\psi'_-,Q'_->.\nonumber \\
\label{cfqcd1vb}
\eea
This is because a change of integration variable from unprimed variable to primed variable does not change the value of the
integration.

The equation
\bea
Q'^c_{\mu \pm}(x)= Q^a_{\mu \pm}(x) +gf^{cba}\omega^a_\pm(x) Q_{\mu \pm}^b(x) + \partial_\mu \omega^c_\pm(x)
\label{thetaav}
\eea
in eq. (\ref{theta}) is valid for infinitesimal transformation ($\omega << 1$) which is obtained from the
finite equation
\bea
T^cQ'^c_{\mu \pm}(x) = U_\pm(x)T^cQ^c_{\mu \pm}(x) U^{-1}_\pm(x)+\frac{1}{ig}[\partial_\mu U_\pm(x)] U^{-1}_\pm(x),~~~~~~~~~~U_\pm(x)=e^{igT^c\omega^c_\pm(x)}.\nonumber \\
\label{ftgrm}
\eea
Simplifying infinite numbers of non-commuting terms we find
\bea
{Q'}_{\mu \pm}^c(x) = [e^{gM_\pm(x)}]_{cb}Q_{\mu \pm}^b(x) ~+ ~[\frac{e^{gM_\pm(x)}-1}{gM_\pm(x)}]_{cb}~[\partial_\mu \omega^b_\pm(x)].
\label{teq}
\eea
where
\bea
M^{cb}_\pm(x)=f^{cba}\omega^a_\pm(x).
\label{mab}
\eea
The fermion field transforms as
\bea
\psi'_\pm(x) =e^{igT^c\omega^c_\pm(x)}\psi_\pm(x).
\label{phg3}
\eea

Under the finite transformation, using eqs. (\ref{teq}), (\ref{phg3}) and \cite{nayakarx} we find
\bea
&& [dQ'_\pm] =[dQ_\pm],~~~~[d{\bar \psi}'_\pm] [d \psi'_\pm ]=[d{\bar \psi}_\pm] [d \psi_\pm ],~~~{F^a}_{\mu \nu}^2[Q'_\pm]={F^a}_{\mu \nu}^2[Q_\pm],~~~~(G_f^c(Q'_\pm))^2 = (\partial_\mu Q^{\mu c}_\pm(x))^2, \nonumber \\
&&{\bar \psi}'_\pm [i\gamma^\mu \partial_\mu -m +gT^c\gamma^\mu Q'^c_{\mu \pm}] \psi'_\pm={\bar \psi}_\pm [i\gamma^\mu \partial_\mu -m +gT^c\gamma^\mu Q^c_{\mu \pm}]\psi_\pm,\nonumber \\
&&{\rm det} [\frac{\delta G_f^c(Q'_\pm)}{\delta \omega^d_\pm}]={\rm det}[\frac{ \delta (\partial_\mu Q^{\mu c}_\pm(x))}{\delta \omega^d_\pm}],~~~~~~~~~~{Q'}_{\mu \pm}^c(x) -A_{\mu \pm}^c(x)= [e^{gM_\pm(x)}]_{cb}Q_{\mu \pm}^b(x).
\label{psa}
\eea
Since we are working in the frozen ghost formalism at initial time
the $<Q_+,\psi_+,{\bar \psi}_+,0|~\rho~|0,{\bar \psi}_-,\psi_-,Q_->$
corresponding to initial density of states in eq. (\ref{nezf}) is gauge invariant by definition.
Hence from eqs. (\ref{teq}) and (\ref{phg3}) we find \cite{nayaka3}
\bea
<Q'_+,\psi'_+,{\bar \psi}'_+,0|~\rho~|0,{\bar \psi}'_-,\psi'_-,Q'_->=<Q_+,\psi_+,{\bar \psi}_+,0|~\rho~|0,{\bar \psi}_-,\psi_-,Q_->.
\label{defn}
\eea

Using eqs. (\ref{psa}) and (\ref{defn}) in eq. (\ref{cfqcd1vb}) we find
\bea
&& <in|Q_{\mu r}^a(x_1)Q_{\nu s}^b(x_2)|in>_A=\int [dQ_+] [dQ_-][d{\bar \psi}_+] [d{\bar \psi}_-] [d \psi_+ ] [d\psi_-]~
~[e^{gM_r(x_1)}]_{ac}Q^c_{\mu r}(x_1)\nonumber \\
&&[e^{gM_s(x_2)}]_{bd}Q^d_{\nu s}(x_2)
{\rm det}(\frac{\delta \partial_\mu Q_+^{\mu c}}{\delta \omega_+^d}){\rm det}(\frac{\delta \partial_\mu Q_-^{\mu c}}{\delta \omega_-^d})
{\rm exp}[i\int d^4x [-\frac{1}{4}{F^c}_{\mu \nu}^2[Q_+]+\frac{1}{4}{F^c}_{\mu \nu}^2[Q_-]\nonumber \\
&&-\frac{1}{2 \alpha} (\partial_\mu Q_+^{\mu c })^2+\frac{1}{2 \alpha}(\partial_\mu Q_-^{\mu c })^2
+{\bar \psi}_+ \dslash [Q_+] \psi_+ -{\bar \psi}_- \dslash [Q_-] \psi_-]]<Q_+,\psi_+,{\bar \psi}_+,0|~\rho~|0,{\bar \psi}_-,\psi_-,Q_->.\nonumber \\
\label{cfq5p1}
\eea
Using the similar technique as above we find
\bea
&&
<in|[e^{gM_r(x_1)}]_{ac}Q_{\mu r}^c(x_1) [e^{gM_s(x_2)}]_{bd}Q_{\nu s}^d(x_2)|in>_A
=\int [dQ_+] [dQ_-][d{\bar \psi}_+] [d{\bar \psi}_-] [d \psi_+ ] [d\psi_-]\nonumber \\
&&Q^a_{\mu r}(x_1)~Q^b_{\nu s}(x_2){\rm det}(\frac{\delta \partial_\mu Q_+^{\mu c}}{\delta \omega_+^d}){\rm det}(\frac{\delta \partial_\mu Q_-^{\mu c}}{\delta \omega_-^d}){\rm exp}[i\int d^4x [-\frac{1}{4}{F^c}_{\mu \nu}^2[Q_+]+\frac{1}{4}{F^c}_{\mu \nu}^2[Q_-]\nonumber \\
&&-\frac{1}{2 \alpha} (\partial_\mu Q_+^{\mu c})^2+\frac{1}{2 \alpha}(\partial_\mu Q_-^{\mu c})^2+{\bar \psi}_+ \dslash [Q_+] \psi_+ -{\bar \psi}_- \dslash [Q_-] \psi_-]]<Q_+,\psi_+,{\bar \psi}_+,0|~\rho~|0,{\bar \psi}_-,\psi_-,Q_->\nonumber \\
\label{cfq5p2}
\eea
where $M^{bc}_r(x)$ is given by eq. (\ref{mab}). From eqs. (\ref{nezf}) and (\ref{cfq5p2}) we find
\bea
<in|Q_{\mu r}^a(x_1)Q_{\nu s}^b(x_2)|in> =<in|[e^{gM_r(x_1)}]_{ac}Q_{\mu r}^c(x_1) [e^{gM_s(x_2)}]_{bd}Q_{\nu s}^d(x_2)|in>_A
\label{finaly}
\eea
which is valid in covariant gauge, in light-cone gauge, in general axial gauges, in general non-covariant gauges and in
general Coulomb gauge etc. \cite{nayakarx}.
Note that the creation operator $a^\dagger$ and annihilation operator $a$ of the quark are related to the quark field
via the equation \cite{mandl}
\bea
&& \psi(x) = \sum_{\rm spin} \sum_p \sqrt{\frac{m}{VE_p}}
 [a(p) u(p) e^{-ip\cdot x} + a^\dagger(p)v(p) e^{ip\cdot x} ]
\eea
where color indices are suppressed. Hence one finds that the quark field $\psi(x)$ depends on the
the creation (annihilation) operator $a^\dagger~(a)$ of the quark but is independent of the creation (annihilation)
operator $a^\dagger_H~(a_H)$ of the hadron. Similarly the gluon field $Q^{\mu a}(x)$ is also independent of the creation
(annihilation) operator $a^\dagger_H~(a_H)$ of the hadron.
Since $a^\dagger_H a_H$ is independent of $\psi(x)$ and $Q^{\mu a}(x)$
one can perform exactly the similar steps of the path integral calculation
as above to find from eq. (\ref{finaly}) that
\bea
<in|Q_{\mu r}^a(x_1)a^\dagger_H a_HQ_{\nu s}^b(x_2)|in> =<in|\Phi_{ac r}(x_1)Q_{\mu r}^c(x_1) a^\dagger_H a_H\Phi_{bd s}(x_2)Q_{\nu s}^d(x_2)|in>_A \nonumber \\
\label{finalz}
\eea
which proves factorization of soft (infrared) divergences at all order in coupling constant in non-equilibrium QCD where
[see eqs. (\ref{mab}) and (\ref{wilt})]
\bea
\Phi_{bc r}(x)=[{\cal P}{\rm exp}[-ig\int_0^{\infty} d\lambda l\cdot { A}_r^a(x+l\lambda)T^{(A)a}]]_{bc},~~~~~~(T^{(A)a})_{bc}=-if^{abc}
\label{wilabf}
\eea
is the non-abelian gauge link or non-abelian phase in the adjoint representation of SU(3),
$l^\mu$ is the light-like four velocity and $r=+,-$ corresponds to the upper and
lower time branches in the closed-time path formalism \cite{keldysh,schw} and $a,b,c=1,2,...,8$ are color indices.

\section{ Gauge Invariant Definition of Gluon to Hadron Fragmentation Function in Non-Equilibrium QCD }

The definition of the gluon to hadron fragmentation function in non-equilibrium QCD in eq. (\ref{gnon}) is not gauge invariant.
To make it gauge invariant and consistent with factorization of infrared divergences we need to incorporate Wilson lines.
The non-abelian gauge link in the adjoint representation of SU(3) is given by the path ordered exponential, see eq. (\ref{wilabf}),
\bea
\Phi_{ab}(x)=[{\cal P}{\rm exp}[-ig\int_0^{\infty} d\lambda l\cdot { A}^c(x+l\lambda)T^{(A)c}]]_{ab},~~~~~~~~~~(T^{(A)c})_{ab}=-if^{abc}
\label{wilabfv}
\eea
which under non-abelian gauge transformation, as given by eq. (\ref{aftgrmpi}), transforms as
\bea
\Phi_{ab}'(x)=U_{ac}(x)\Phi_{cb}(x)
\label{ty}
\eea
where
\bea
U_{ab}(x)=[e^{igT^{(A)c}\omega^c(x)}]_{ab}
\label{tyadj}
\eea
and from eq. (\ref{jprtpi}) we find
\bea
Q'^a_{\mu r}(x)=U_{ab}(x)Q_{\mu r}^b(x).
\label{jpr}
\eea

Hence by extending the path integral formulation of quark fragmentation function in non-equilibrium QCD \cite{nayaka3} to gluon
fragmentation function in non-equilibrium QCD we find from eqs. (\ref{finalz}), (\ref{wilabf}),
(\ref{gnon}), (\ref{nezf}), (\ref{ty}) and (\ref{jpr})
that the gauge invariant definition of the gluon to hadron fragmentation function in non-equilibrium QCD
which is obtained from the single gluon {\it in}-state $|g>$ and is consistent with
factorization theorem in non-equilibrium QCD is given by
\bea
&& D_{H/g}(z,P_T)
= \frac{k^+}{16z~[1+f_g(k^+,k_T)]} \int dx^- \frac{d^{d-2}x_T}{(2\pi)^{d-1}}  e^{i{k}^+ x^- + i {P}_T \cdot x_T/z} \nonumber \\
&&<in| Q^{\mu a}(x^-,x_T) \Phi[x^-,x_T]a^\dagger_H(P^+,0_T)  a_H(P^+,0_T)  \Phi[0]Q_\mu^a(0) |in>
\label{gnff}
\eea
which reproduces eq. (\ref{gnf}) which is valid in covariant gauge, in light-cone gauge, in general axial gauges,
in general non-covariant gauges and in general Coulomb gauge etc. where the path ordered exponential
$\Phi_{ab}[x ]$ is given by eq. (\ref{wilg}).

This completes the derivation of gluon to hadron fragmentation function in non-equilibrium QCD.

\section{ Conclusions }
Theoretical understanding of the observed jet quenching measurements at RHIC and LHC
is challenging in QCD because it requires understanding of parton to hadron fragmentation
function in non-equilibrium QCD. In this paper, by using closed-time path integral formalism,
we have derived the gauge invariant definition of the gluon to hadron fragmentation function
in non-equilibrium QCD which is consistent with factorization theorem in non-equilibrium
QCD from first principles.

\end{document}